# Optimal-speed algorithms for localization of random pulsed point sources generating super short pulses

A.L.Reznik, A.A.Soloviev, A.V.Torgov

Institute of Automation and Electrometry, Siberian Branch, Russian Academy of Sciences, Academician Koptyug ave. 1, Novosibirsk, Russia, 630090

**Corresponding author: A.L.Reznik**
E-mail: reznik@iae.nsk.su

*Abstract.* The time-optimal technique of spatial localization of the random pulsed-point source that has the uniform distribution density on search interval and indicating itself by generation of the instant impulses (delta functions) at random time points is developed. Localization is carried out by means of the receiver having view window freely reconstructed in time. The created algorithms are generalized to the search carried out by system consisting of several receivers.

*Introduction*

Usually one of two types of a location – passive or the active is applied to search for unknown signal sources [1-3]. The passive location is based on reception of own radiation of the object. In case of the active location the search system radiates own probing signal and receives its reflection from the object. Depending on parameters of the received signal coordinates (characteristic) of the target are defined. In this paper we consider algorithms of time-optimal search for Poisson sources indicating themselves by generation pulses at random timepoints. The major feature which is essentially distinguishing our research from the passive and active location methods mentioned above is that the searched source becomes itself apparent by means of generation of the *infinitely short* (*instantaneous*) delta-pulses. Time-optimal search algorithm should, generally, meet one of two requirements: either to minimize the total search effort necessary to detect the object; or to maximize the full detection probability in the presence of a limited search effort. We understand the pulsed-point source as an object with negligible small angular dimensions (a mathematical point) that placed randomly on the $x$ axis with a priori probability density function $f(x)$ and emitting infinitely short pulses ($\delta$-functions) with a Poisson intensity $\lambda$. Thus, the time interval between pulses is random variable $t$ with an exponential density function $h(t)=\lambda exp(-\lambda t)$. The search for the object is performed by a recording device, whose «viewing window» is capable to be reconstructed in any way in time. The pulse is fixed if the active object produced the pulse is in the "viewing window" of the recording device. Otherwise, the pulse is considered to be missed. As the pulse is registered, the window narrows, and the source position becomes more precise. It is required to search for the source during the minimum (statistically) time with an $\varepsilon$ accuracy.

In §1 the general problem definition of localization of a random pulsed source is given and the complexity of its analytical decision in case of source's arbitrary distribution on a search interval is shown. In the same section the strategies of search, optimum in certain classes of search algorithms are calculated. All tasks considered in §1 belong to localization of the only source of impulses by means of the only receiver.

In §2 the problem of consecutive localization of several pulsed sources is considered. It is supposed that localization of each of them is carried out by means of one receiver, but, unlike §1, it is considered that all sources are uniformly distributed on a search interval. For construction optimum search strategy are used not only traditional numerical algorithms, but also program algorithms of carrying out symbolic calculations and analytical transformations which are seldom used in scientific practice.

In §3 the problem of creation of the strategy of localization of a random pulsed source, optimum on time, when the reception system consists of several receivers is solved. As well as §2, it is supposed that the unknown source has uniform distribution on a search interval.

*1. Programs and algorithms for optimal-speed search of single pulsed point object*

(a) One-step search algorithms. Introducing the binary function $u(x,t)$ describing the viewing window at time $t$, we obtain the relation for the average time from the beginning of the search to recording the first pulse [3]:

$$\langle \tau \rangle = \lambda \int_0^\infty dt \int_0^\infty dx \left[ t f(x) u(x,t) \exp\left(-\lambda \int_0^\infty u(x,\xi) d\xi\right) \right]. \tag{1}$$

For the random a priori distribution of the source on the axis $x$, the construction even one-step (ending after registering the first pulse) optimal-time search procedure causes considerable difficulties. In one-step *periodic* search algorithms, the relative load $\varphi(x)$ per point $x$ (i.e., the relative time the point is present in the viewing window) remains constant throughout the search time. With this approach, the task is to find the function that minimizes the average search time

$$\langle \tau \rangle = \frac{1}{\lambda} \int \frac{f(x)}{\varphi(x)} dx \tag{2}$$

provided that

$$\int \varphi(x) dx = \varepsilon, \tag{2a}$$

$$0 \leq \varphi(x) \leq 1. \tag{2b}$$

Using the method of undetermined Lagrange multipliers, we search for the function $\varphi(x)$ that minimizes the expression

$$\int \left[ \frac{f(x)}{\varphi(x)} + \mu \varphi(x) \right] dx.$$

Further, differentiating with respect to $\varphi$ and taking into account the constraint (2a), we get

$$\varphi(x) = \varepsilon \sqrt{f(x)} \bigg/ \int \sqrt{f(x)} dx. \tag{3}$$

If at the same time the condition (2b) is not violated, then the function (3) is a solution of the formulated extreme problem. If there are such areas of $x$ where the solution $\varphi(x)>1$, then inside of these areas it is necessary to set $\varphi(x)=1$, and for the remaining points it is required to recalculate the undefined multiplier $\mu$ under already changed conditions (2) and (2a). After this, as an optimal search strategy, any binary function $u(x,t)$ can be chosen that satisfies the relations

$$\int u(x,t) dx = \varepsilon; \qquad \int u(x,\xi) d\xi = \varphi(x) t.$$

In the general case, the construction of the optimal (not necessarily periodic) one-step search algorithm for an unknown Poisson source is connected with finding such a function $\varphi(x,t)$ – the relative load on the point $x$ at a time $t$ that minimizes the average search time

$$\langle \tau \rangle = \int dt \int dx f(x) \exp\left(-\lambda \int_0^t \varphi(x,\xi)d\xi\right) \qquad (4)$$

provided that

$$\int \varphi(x,t)dx = \varepsilon \text{ for any } t, \qquad (4a)$$

$$0 \leq \varphi(x,t) \leq 1. \qquad (4b)$$

To simplify further calculations, we introduce a function $\alpha(x,t) = \int_0^t \varphi(x,\xi)d\xi$ corresponding to the total time spent by the point $x$ in the viewing window from the beginning of the search to time $t$. To take into account the constraints (4a) and (4b), we introduce the undetermined Lagrange multiplier $\mu(t)$. Then the problem of finding the optimal strategy reduces to finding the function $\alpha(x,t)$ that minimizes the functional

$$\int dt \int dx \left[\exp(-\lambda \alpha(x,t))f(x) + \mu(t)\alpha(x,t)\right] \qquad (5)$$

provided that

$$\int_{-\infty}^{\infty} \alpha(x,t)dx = \varepsilon t, \qquad (5a)$$

$$0 \leq \alpha(x,t) \leq t. \qquad (5b)$$

The solution of this variational problem is the function

$$\alpha(x,t) = \begin{cases} 0, \dfrac{1}{\lambda}\ln\dfrac{\lambda f(x)}{\mu(t)} < 0; \\ \dfrac{1}{\lambda}\ln\dfrac{\lambda f(x)}{\mu(t)}, 0 \leq \dfrac{1}{\lambda}\ln\dfrac{\lambda f(x)}{\mu(t)} \leq t; \\ t, t < \dfrac{1}{\lambda}\ln\dfrac{\lambda f(x)}{\mu(t)}, \end{cases} \qquad (6)$$

where $\mu(t)$ is determined from the relation (5a). The optimal search strategy $u(x,t)$ must belong to the class of binary functions. It is set by the equations

$$\int_0^t u(x,\xi)d\xi = \alpha(x,t); \qquad \int u(x,t)dx = \varepsilon.$$

The use of optimal search algorithms in practice faces certain difficulties. The fact is both proposed algorithms of optimal one-step search, when a priori density function of the signal source differs from the constant, can not be physically realized (as a rule) by moving an 1-connected viewing window with a width $\varepsilon$. Therefore, in real search procedures, one-step procedure is advisable to do in according to the following scheme.

Preliminarily, the interval $(0,L)$ is divided into $\varepsilon$-step discretes, and a priori given density $f(x)$ is "stepwise" approximated on each of them. In this case, $\varepsilon$ is considered to be a sufficiently small so that the variation of the function $f(x)$ within one discrete can be neglected. Search (in accordance with the optimal one-step detection procedure) should begin only with the highest "peak" density inspection, then after a time $t_1$ the window is alternately set to the areas under the two highest "peaks", then after time $t_2$, further inspection is done for three sections, and so on. All the moments $t_i$ are exactly determined by the described finding method for optimal function $\alpha(x,t)$.

It should be noted that discussed search plan assumes that the intensity of the source $\lambda$ is known in advance. If such a priori information is not available, a periodic procedure can be recommended that does not depend on this intensity. In accordance with it, the integrals of the density $f(x)$ in each of the discrete must be calculated. If there are $m$ discretes, and its squares-integrals are $P_1, P_2, \ldots, P_m$, then the viewing window should cyclically "run through" all the discretes with relative load $\beta_i = \sqrt{P_i} / \sum_{j=1}^{m} \sqrt{P_j}$ ($j=1,\ldots,m$), respectively. These values $\beta_i$ are easily obtained if we again apply the method of undetermined Lagrange multipliers to minimize the average search time

$$\langle \tau \rangle = (1/\lambda)(P_1/\beta_1 + P_2/\beta_2 + \ldots + P_m/\beta_m) \tag{7}$$

provided that

$$\beta_1 + \ldots + \beta_m = 1. \tag{8}$$

(b) Multistep search algorithms. If we are not limited to one-step procedures, but consider the search algorithm as a multi-step process (that ends after $n$-th pulse registration), then the optimal strategy should deliver a minimum to the functional

$$\langle \tau \rangle = \sum_{k=1}^{n} \lambda^k \int_{0}^{\infty} dx f(x) \int \ldots \int t_k \times \left\{ \prod_{l=1}^{k} \left[ dt_l u_l(x, \sum_{m=1}^{l} t_m, t_1, \ldots, t_{l-1}) \exp(-\lambda \int_{\sum_{m=1}^{l-1} t_m}^{\sum_{m=1}^{l} t_m} u_l(x, \xi, t_1, \ldots, t_{l-1}) d\xi) \right] \right\}, \tag{9}$$

provided that

$$\int u_n(x, t, t_1, \ldots, t_{n-1}) dx = \varepsilon. \tag{10}$$

Here $U_i(x,t,t_1,\ldots t_{i-1})$ is the search strategy at the $i$-th step provided that the intervals between the first ($i$-1) pulses were $t_1, t_2, \ldots, t_{i-1}$ respectively. In the general case, to find the optimal strategy $u(x,t)$ that minimizes the functional (9) is not possible. At the same time, for an important special case, $f(x)=const$, the analytic solution is rather simple. Let

$$f(x) = \begin{cases} \frac{1}{L}, & x \in (0, L) \\ 0, & x \notin (0, L), \end{cases}$$

i.e. there is no a priori information about placement of the source within the interval (0,$L$). Obviously, in the first step, the search load should be equally distributed between all the points $x \in (0, L)$. It is possible to make such a load, for example, by television scanning of the whole interval (0,$L$) by the aperture with width $l_1$ (to avoid edge effects, we consider the end of the interval closed to its beginning, so, a circle with length $L$ is scanned instead of the interval). When registering a pulse, the search continues inside the window with width $l_1$ using another aperture with width $l_2$. If we discuss an $n$-step search, then at the last step scanning it is carried out by the aperture with width $\varepsilon$ (this is dictated by the conditions of the task). Then the average search time is

$$\langle \tau \rangle = (1/\lambda)(L/l_1 + l_1/l_2 + \ldots + l_{n-1}/\varepsilon). \tag{11}$$

For a fixed $n$, it's possible to find optimal values that minimize the expression (11):

$$L/l_1 = l_1/l_2 = \ldots = l_{n-1}/\varepsilon = (L/\varepsilon)^{1/n}. \tag{12}$$

Then the average time of the optimal $n$-step search is

$$\langle \tau_n \rangle_{opt} = (n/\lambda)(L/\varepsilon)^{1/n}. \tag{13}$$

Now (from the expression (13)) we can find the optimal number of steps $n$ minimizes the average search time. Since the function $xa^{1/x}$ for $a>1$ has only one minimum point ($x=\ln a$), the optimal value $n_{opt}$ is always either *entier*($\ln(L/\varepsilon)$) or *entier*($\ln(L/\varepsilon)$)+1. When $L/\varepsilon \to \infty$, it is possible to consider $n_{opt} \cong \ln(L/\varepsilon)$. Therefore, we have asymptotic relations:

$$L/l_1 = l_1/l_2 = ... = l_{n-1}/\varepsilon = e. \tag{14}$$

$$\langle \tau_n \rangle_{opt} = (n_{opt}/\lambda)(L/\varepsilon)^{1/n_{opt}} = (e/\lambda)\ln(L/\varepsilon). \tag{15}$$

So, a multi-step procedure brings a gain in comparison with a one-step search (for a one-step search in the case under consideration, the average detection time is $\langle \tau_1 \rangle_{opt} = L/\lambda\varepsilon$), and this gain increases unlimitedly with the ratio $L/\varepsilon$ increasing. Now we can compare the constructed optimal search procedure with some simplified algorithms. For example, if the search is organized according to the principle of a dichotomy, then the average time of the object search is

$$\langle \tau \rangle_2 = (2/\lambda)\log_2(L/\varepsilon) = (2/\lambda \ln 2)\ln(L/\varepsilon). \tag{16}$$

Dichotomous search has (in comparison with the optimal procedure) a small ($\approx 6\%$) loss in time. Trichotomic search is even closer to the optimal procedure: initial interval (0,$L$) is divided into three subintervals, then the subinterval where impulse is fixed, in turn, is also divided into three subintervals, etc. This procedure loses to the optimal procedure only

$$(3/\ln 3 - e)/e \approx 0{,}4\%.$$

It is natural to expect that in the case of arbitrariest a priori distribution $f(x)$ the multi-step search procedure (in comparison with one-step search) can bring a significant gain in time, especially for large values of $L/\varepsilon$. Since minimization of the functional (9) under the constraint (10) in each concrete case is very difficult problem, multi-step *periodic* search procedure seems to be more real from the practical realization point of view. In the first step, the interval (0,$L$) is divided into three parts (three parts are chosen because for a uniform distribution $f(x)$ this procedure is closest to the optimal one). Then the values $P_1, P_2, P_3$ are calculated

$$P_1 = \int_0^{L/3} f(x)dx; \qquad P_2 = \int_{L/3}^{2L/3} f(x)dx; \qquad P_3 = \int_{2L/3}^{L} f(x)dx.$$

For any time interval $\Delta t$, viewing window with width $L/3$ will periodically "run through" all three sections in such a way that $\Delta t_1 + \Delta t_2 + \Delta t_3 = \Delta t$, where $\Delta t_i / \Delta t = \beta_i$ is the relative viewing window presence time on each of the subintervals (0,$L/3$), ($L/3$,$2L/3$), ($2L/3$,$L$). When the pulse is registered, the search procedure continues similarly on the subinterval (segment) where the pulse is fixed (i.e., this segment is divided again into three parts, the coefficients $\beta_1, \beta_2, \beta_3$ are recalculated, etc.). At the first step their values are equal $\sqrt{P_i} / \sum_{j=1}^{3} \sqrt{P_j}$ ($i=1,2,3$). The universality of the proposed procedure is also in the fact that its realization does not require a priori information on the source intensity.

## *2. Symbolic-analytic and numerical algorithms in the problem of optimal multipurpose search*

In.§2 we investigate analytic and numerical algorithms for optimal multipurpose search. The main problem considered here, can be formulated as follows. There are $n$ point objects on the segment (0,$L$). There is no a priori information about their location, so we can assume that $n$ points are uniformly distributed on the segment (0,$L$). Each object at random moments of time produces short pulses (delta functions), the pauses between them have an exponential distribution with the parameter $\lambda$. As before, the pulse is fixed if the active object initiated the pulse is in the viewing window of the recording device. It is necessary to localize all sources with $\varepsilon$ accuracy in the minimum time. Generally speaking, the construction of an optimal multipurpose search strategy can be reduced to a sequential procedure, when at the first step the localization of one of the $n$ sources is carried out with required accuracy, then one of the ($n$-1) remaining sources is searched for, etc. Two different approaches are possible here: either all "previous steps" accumulated information is used, or there is no such an accumulation. The use of previously accumulated information makes search algorithms and the scanning system itself more complicated, because in this case it has to have an opportunity to store and to process intermediate data.

The algorithms considered below are constructed under the assumption that the recording equipment has no memory, so the problem described above reduces to finding the optimal search strategy for one of $n$ point objects with an accuracy $\varepsilon$. The search procedure for each of the $n$ steps, in turn, can also be considered as a multi-step process. Since initially there is no a priori information about the location of $n$ sources within the segment (0,$L$), in the first step the search effort should be equally distributed between all the points $x \in (0, L)$. You can implement such a scheme, for example, by scanning the whole segment by aperture of some width $l_1$. Then (after registering the pulse), the search continues inside the selected window $l_1$ by another aperture of width $l_2$, etc. At the last step (the number of such steps depends on $n$, $\varepsilon$, $L$) to ensure required localization accuracy, the search is carried out by the aperture of width $\varepsilon$. The task now is to determine the optimal number of scanning steps for the given values $n$, $\varepsilon$ and $L$ the width of the scanning viewing window for each of them.

After finding one of the objects, the value $n$ is reduced by one, and the search is repeated for $n=n-1$, etc. We introduce the function $g(x)$ – the density distribution of the aperture center at the time of the first pulse registration. If we assume that $n$ sources are located at points $x_1, x_2, \ldots x_n$, and the scanning is carried out by the aperture of width $l$, then

$$g(x) = \frac{1}{nl} \sum_{i=1}^{n} 1[x - (x_i - 0.5l)] \times 1[(x_i + 0.5l) - x], \qquad (17)$$

where $1[x] = \begin{cases} 0, & x \leq 0, \\ 1, & x > 0 \end{cases}$ – unit step function (Heaviside step function).

Further in the text, we'll need knowledge of the number of objects distribution in the aperture at the moment of recording the pulse. To find this distribution, it's required to calculate $P_n(k,l)$ – the probability that will be exactly $k$ ($k=1,2,\ldots,n$) sources in the aperture at the time of pulse registration, including the object generated the pulse. Since the realization of the scheme with uniform search effort leads to undesirable edge effects, then in order to make the subsequent presentation stricter, we assume that search interval is the circle with circumference $L$. Moreover, without loss of generality we assume that $L=1$, and $0 < l \leq 1$.

Let the objects be located at points $x_1, x_2, \ldots x_n$. Assuming the point $x_1$ to be the origin, we rank (put in order) all other sources moving from the point $x_1=0$ clockwise. The probability $P_n(k,l)$ consists of the probabilities that, at the moment of pulse fixation, either the aperture contains points $x_1, x_2, \ldots x_k$ , or $x_2, x_3, \ldots x_{k+1}$ and so on. The number of such combinations is $n$. If we assume that at the moment of pulse registration there are exactly $k$ objects in the aperture, then taking into account (17)

$$P_n(k,l) = \frac{(n-1)! \times k \times 2}{l} \int \ldots \int_D \min\{l - x_k; x_{k+1} - x_k; x_{k+1} - l + 1\} dx_2 \ldots dx_n, \qquad (18)$$

where the area $D$ is given by the system of inequalities:

$$\begin{cases} 0 < x_2 < x_3 < ... < x_n < 1, \\ x_{k+1} - x_k \leq 1 - x_n, \\ x_{k+1} + (1 - x_n) > l, \\ x_k < l. \end{cases} \quad (19)$$

The doubling of the coefficient in relation (18) is taking place because inequality $x_{k+1} - x_k \leq 1 - x_n$ is included in the constraint system (19), the geometric meaning of which lies in the fact that one of the intervals (formed as a result of random throwings of $n$ points into a circle) is greater than the other interval. Since the probability of such an event is 1/2, then the cofactor 2 appears in the expression $P_n(k,l)$ (note that in the case $k=n$ the inequality $x_{k+1} - x_k \leq 1 - x_n$ turns into the identity inequality, so it is simply excluded from the system (19), and doubling cofactor removes from (18)).

The integral (18) over the area (19) can be represented as the sum of three integrals that correspond (in form) to the expressions for computer analytic calculations described in [4]:

$$P_n(k,l) = \frac{2k(n-1)!}{l} \left\{ \int ... \int_{D_1} (l - x_k) dx_2 ... dx_n + \int ... \int_{D_2} (x_{k+1} - x_k) dx_2 ... dx_n + \int ... \int_{D_3} (x_{k+1} - l + 1 - x_n) dx_2 ... dx_n \right\}. \quad (20)$$

Here, the following systems of linear inequalities correspond to regions $D_1$, $D_2$, $D_3$ with one free parameter $l$:

$$D_1 = \begin{cases} x_1 = 0 < x_2 < x_3 < ... x_n < 1 = x_{n+1}, \\ x_{k+1} - x_k < 1 - x_n, \\ x_{k+1} + (1 - x_n) > l, \\ x_k < l, \\ x_{k+1} - l > 0, \\ x_k + x_{k+1} - x_n + 1 - 2l > 0, \end{cases} \quad (21)$$

$$D_2 = \begin{cases} x_1 = 0 < x_2 < x_3 < ... x_n < 1 = x_{n+1}, \\ x_{k+1} - x_k < 1 - x_n, \\ x_{k+1} + (1 - x_n) > l, \\ x_k < l, \\ l - x_{k+1} > 0, \\ x_k - x_n + 1 - l > 0, \end{cases} \quad (22)$$

$$D_3 = \begin{cases} x_1 = 0 < x_2 < x_3 < ... x_n < 1 = x_{n+1}, \\ x_{k+1} - x_k < 1 - x_n, \\ x_{k+1} + (1 - x_n) > l, \\ x_k < l, \\ -x_k - x_{k+1} + x_n - 1 + 2l > 0, \\ -x_k + x_n - 1 + l > 0. \end{cases} \quad (23)$$

Computer calculations carried out for fixed values $n$ and $k$ have shown that in the general case

$$P_n(k,l) = \binom{n-1}{k-1} l^{k-1} (1-l)^{n-k}. \quad (24)$$

In fact, this means that we are dealing with a "shifted" Bernoulli scheme: $(n-1)$ independent tests are carried out with the probability of "success" $p=l$ and the probability of "failure" $q=1-l$, and formula (24) corresponds to the probability that there will be exactly $(k-1)$ "successes". Now, when a program-calculated expression for probability $P_n(k,l)$ is known, we can also construct a proof that computer analytic calculations are correct.

Since the scanning of the unit circle (i.e. circle whose circumference is 1) by arc-aperture of length $l$ creates a uniform search effort over all points that form circumference of the circle, the probability the initiator of the pulse is a fixed source $x_i$ ($i=1,2,...,n$) is the same for all objects and is equal to $1/n$. Therefore, by the formula of total probability

$$P_n(k,l) = \sum_{i=1}^{n} P(A_i) P_n(k,l/A_i) = \frac{1}{n} \sum_{i=1}^{n} P_n(k,l/A_i) = P_n(k,l/A_1), \quad (25)$$

where the event $A_i$ means that the initiator of the registered impulse was the object $x_i$. The conditional probability $P_n(k,l/A_1)$ can be interpreted as follows: all sources except the first are "locked", and at the moment of pulse registering the number of objects in the aperture, including the first one, is counted. Since the remaining ($n-1$) sources have a uniform distribution on the circle, the probability for any of them to get into the aperture is $l$, and the probability that the aperture will contain exactly ($k-1$) objects (except the first) is just described by the relation (24).

From (24), it can be shown that if we sequentially scan a unit circle by the aperture $l_1$ ($l_1<1$) and then (after pulse registration) the circle of circumference $l_1$ (this circle is the arc $l_1$ converted) is scanned by another aperture $l_2$ ($l_2<l_1$), then the probability that at the moment of second pulse registration, there will be exactly $k$ objects in the aperture is the same as the probability that the aperture will contain exactly $k$ objects if initial unit circle is scanned at once by the aperture of width $l_2$. In fact, the probability that $k$ objects will get into the aperture of width $l_2$ during a double scan is

$$\sum_{i=k}^{n} P_n(i,l_1) P_i(k,\frac{l_2}{l_1}) = \sum_{i=k}^{n} \binom{n-1}{i-1} l_1^{i-1}(1-l_1)^{n-i} \binom{i-1}{k-1} \left(\frac{l_2}{l_1}\right)^{k-1} \left(1-\frac{l_2}{l_1}\right)^{i-k} =$$
$$= \binom{n-1}{k-1} l_2^{k-1} \sum_{i=k}^{n} \binom{n-k}{n-i} (l_1-l_2)^{i-k}(1-l_1)^{n-i} = \binom{n-1}{i-1} l_2^{k-1} \sum_{i=0}^{n-k} \binom{n-k}{i} (l_1-l_2)^{i}(1-l_1)^{n-k-i} = \quad (26)$$
$$= \binom{n-1}{i-1} l_2^{k-1}(1-l_2)^{n-k},$$

So, it does not depend on what aperture was used for the first scan. Using the relations (24) and (26), we find the average time of the objects $n$ localization with a given accuracy $\varepsilon$, assuming that the search is carried out with the help of several steps by consistently narrowed apertures $l_1,l_2,...,l_{m-1},\varepsilon$. To do this, we first calculate the average search time at the $i$-th step:

$$\langle \tau_i \rangle = \sum_{k=1}^{n} \binom{n-1}{k-1} l_{i-1}^{k-1}(1-l_{i-1})^{n-k} l_{i-1}/(l_i k \lambda) =$$
$$= (l_{i-1}/(\lambda l_i)) \sum_{k=1}^{n} \binom{n}{k} l_{i-1}^{k}(1-l_{i-1})^{n-k}/(n l_{i-1}) = \frac{1-(1-l_{i-1})^n}{n \lambda l_i}. \quad (27)$$

The average search time for one of the sources $n$ is the sum of the average search times at each step, therefore

$$\langle \tau \rangle = \frac{1}{n\lambda} \sum_{i=1}^{m} (1-(1-l_{i-1})^n)/l_i, \quad (28)$$

here $l_0=1$; $l_m=\varepsilon$. The problem of optimal search has now reduced to finding $m,l_1,l_2,...,l_{m-1}$ that minimize expression (28). For a fixed value $m$, the optimal values $l_1,l_2,...,l_{m-1}$ can be calculated from the system of equations

$$l_{i+1} = \frac{n l_i^2 (1-l_i)^{n-1}}{1-(1-l_{i-1})^n}, \quad (i=1,2,...,m-1) \quad (29)$$

that is obtained by equating all the partial derivatives of the expression (28) to zero. The system of equations (29) was solved on computer for fixed values $m$. The optimal $m$ values (i.e. the number of scanning steps) and the corresponding values $l_1, l_2, \ldots, l_{m-1}$ when the average localization time (28) reaches a minimum were calculated for different values $n$ and $\varepsilon/L$. These data are summarized in Table 1. We can see from this table that at high requirements to localization accuracy (when $\varepsilon/L \ll 1$), the average time of search of the first of objects is proportional to $\ln(\varepsilon/L)$.

**Table 1. Parameters of the time-optimal search. Initial data: n - number of impulse objects; L is the length of the segment on which the impulse objects are located; ε - required accuracy of localization; λ - Poisson generation intensity of each pulsed objects. Results of the program calculation: m - the number of scanning steps for the localization of the first object (from n); $l_i$ - the size of the scanning aperture at the i-th step (i = 1,....m); $\langle \tau \rangle$ - the average localization time of the first object.**

| $\varepsilon/L$ | $m, l, T$ \ $n$ | 2 | 3 | 5 | 10 | 30 | 50 |
|---|---|---|---|---|---|---|---|
| $10^{-1}$ | $m$ | 2 | 2 | 1 | 1 | 1 | 1 |
|  | $l_1/L$ | 0.26 | 0.24 | 0.1 | 0.1 | 0.1 | 0.1 |
|  | $l_2/L$ | 0.1 | 0.1 | - | - | - | - |
|  | $\lambda \langle \tau \rangle$ | 4.19 | 3.26 | 2.0 | 1.0 | 0.33 | 0.2 |
| $10^{-2}$ | $m$ | 4 | 4 | 3 | 3 | 1 | 1 |
|  | $l_1/L$ | 0.23 | 0.19 | 0.09 | 0.07 | 0.01 | 0.01 |
|  | $l_2/L$ | 0.08 | 0.07 | 0.03 | 0.03 | - | - |
|  | $l_3/L$ | 0.03 | 0.03 | 0.01 | 0.01 | - | - |
|  | $l_4/L$ | 0.01 | 0.01 | - | - | - | - |
|  | $\lambda \langle \tau \rangle$ | 10.22 | 9.02 | 7.55 | 5.73 | 3.33 | 2.0 |
| $10^{-3}$ | $m$ | 6 | 6 | 6 | 5 | 4 | 3 |
|  | $l_1/L$ | 0.21 | 0.16 | 0.12 | 0.06 | 0.02 | 0.01 |
|  | $l_2/L$ | 0.07 | 0.06 | 0.043 | 0.02 | 0.007 | 0.003 |
|  | $l_3/L$ | 0.024 | 0.02 | 0.016 | 0.007 | 0.003 | 0.001 |
|  | $l_4/L$ | 0.008 | 0.007 | 0.006 | 0.003 | 0.001 | - |
|  | $l_5/L$ | 0.003 | 0.003 | 0.003 | 0.001 | - | - |
|  | $l_6/L$ | 0.001 | 0.001 | 0.001 | - | - | - |
|  | $\lambda \langle \tau \rangle$ | 16.48 | 15.22 | 13.76 | 11.76 | 8.77 | 7.42 |

|  | $m$ | 9 | 8 | 8 | 7 | 6 | 6 |
|---|---|---|---|---|---|---|---|
| $10^{-4}$ | $l_1/L$ | 0.24 | 0.15 | 0.11 | 0.05 | 0.018 | 0.013 |
|  | $l_2/L$ | 0.09 | 0.05 | 0.04 | 0.017 | 0.006 | 0.005 |
|  | $l_3/L$ | 0.03 | 0.02 | 0.014 | 0.006 | 0.002 | 0.0017 |
|  | $l_4/L$ | 0.01 | 0.006 | 0.005 | 0.002 | 0.0008 | 0.0007 |
|  | $l_5/L$ | 0.005 | 0.002 | 0.002 | 0.0008 | 0.0003 | 0.0003 |
|  | $l_6/L$ | 0.002 | 0.0008 | 0.0007 | 0.0003 | 0.0001 | 0.0001 |
|  | $l_7/L$ | 0.0007 | 0.0003 | 0.0003 | 0.0001 | - | - |
|  | $l_8/L$ | 0.0003 | 0.0001 | 0.0001 | - | - | - |
|  | $l_9/L$ | 0.0001 | - | - | - | - | - |
|  | $\lambda\langle\tau\rangle$ | 22.74 | 21.48 | 19.97 | 18.0 | 14.96 | 13.6 |

## 3. Systems with multiple receivers: optimal-speed algorithms for random pulsed point sources search

Problems and algorithms for the optimal search of random sources that will be discussed in this paragraph arise in many scientific and technical fields, in particular, in classical disciplines as reliability theory and mathematical communication theory. Similar studies are necessary for the development of methods for tech troubleshooting, appearing in a form of the alternating equipment failure; in astrophysics these problems are encountered in the search of pulsating radiation sources; in modern sections of computational mathematics these methods are required to develop algorithms for detecting low-contrast and small-sized objects on aerospace images, and, for example, in signal theory, the same methods are used to estimate the reliability of random fields and point images registration.

In §1 and §2 optimal search algorithms for pulsed point sources were described, which assumed the using of a single receiver with a tunable viewing window. In the presence of several receiving devices, the average localization time can be greatly reduced. The purpose of this paragraph 1.3 is to construct the time-optimal (in statistical terms) localization algorithms for randomly located pulsed point source that take into account the number of receivers used and provide the required localization accuracy. Within this study, it will be assumed that a priori information about the probable location of a random pulsed object inside of the search interval $(0,L)$ is absent, i.e. that the probability density of an unknown source on the $x$ axis is given by the function

$$f(x) = \begin{cases} \dfrac{1}{L}, & x \in (0,L) \\ 0, & x \notin (0,L) \end{cases}.$$

The first question we need to answer is: "What accuracy of localization can be achieved with a single-tact search procedure that is carried out with the help of $n$ receiving devices and ends at the time of first pulse generation by source?"

We pay special attention to the fact that, according to the conditions of the problem, the search procedure ends not at the moment of first pulse registration by the receiving system, but at the moment of random source's first pulse generation. This is not the same, because in principle the receiving system can work in "missed-some-pulses" mode (not registering all the generated pulses). Thus, in single-tact search, it is required that the system doesn't miss any impulse. Obviously, in this case, the aggregate viewing window of all $n$ receivers has to overlap the entire search interval $(0,L)$ at any moment. Let $N$ be the number of elementary segments which the interval $(0,L)$ should be divided into (pulsed object is located in one of such a segments). Generally speaking, it would be possible, for example, to put $N$ equal to the number of receivers $n$ and, dividing the search interval into $n$ equal parts, assign to each of $N=n$ segments theirs own available receiver. In such a search procedure, there is no problem to "link" the source to desired segment, since the detected pulse is always fixed by only one receiver. But, unfortunately, such a simplified algorithm has an extremely low accuracy of localization $\varepsilon = \dfrac{L}{n}$, and therefore it is very far from optimal.

(a) <u>Optimal single-tact search procedure.</u> The construction of a single-tact search procedure for $n$ receiving devices, which really minimizes the localization error, will be carried out as follows. Each of the $n$ receivers is specified with the help of Table 2, the $i$-th line of which describes the monitoring zone of the receiver with the number $i$. Binary variables $x_{ij}, i = \overline{1,n}; j = \overline{1,N}$ forming the table (note that the optimal value of the parameter $N$ has yet to be determined) will have the values 0 or 1 according to the following rule:
– if $x_{ij}=0$, segment $j$ does not enter the observation zone of receiver $i$;
– if $x_{ij}=1$, the segment $j$ enters the observation zone of the receiver $i$.

**Table 2. A two-dimensional array describing the observation zone for each of $n$ receiving system's devices in a single-tact search procedure.**

| RECEIVERS | Segments | | | | |
|---|---|---|---|---|---|
| | 1 | 2 | 3 | … | N |
| 1-th receiver | $x_{11}$ | $x_{12}$ | $x_{13}$ | | $x_{1N}$ |
| 2-th receiver | $x_{21}$ | $x_{22}$ | $x_{23}$ | | $x_{2N}$ |
| 3-th receiver | $x_{31}$ | $x_{32}$ | $x_{33}$ | | $x_{3N}$ |
| ⋮ | ⋮ | ⋮ | ⋮ | ⋮ | ⋮ |
| $i$-th receiver | $x_{i1}$ | $x_{i2}$ | $x_{i3}$ | | $x_{iN}$ |
| | ⋮ | ⋮ | ⋮ | ⋮ | ⋮ |
| $n$-th receiver | $x_{n1}$ | $x_{n2}$ | $x_{n3}$ | | $x_{nN}$ |

Thus, the row vector $\mathbf{x}_i=(x_{i1}, x_{i2}, …, x_{iN})$ uniquely defines the receiver $i$, and the whole array $(x_{ij})$ uniquely fully describes the receiving system. The dynamic state of this system is characterized by the column vector $\mathbf{r}=(r_1,r_2,…,r_n)^T$, in which all binary (i.e., possessing the value 0 or 1) variables $r_i$, $i = \overline{1,n}$ are equal to zero throughout the time interval from the beginning of the search to pulse's registration moment. When the pulse is fixed, each of the variables $r_i$ goes (or not) into the state $r_i=1$, depending on whether the received pulse was fixed by the $i$-th receiver or not. The task is: by means of this changed column vector $\mathbf{r}$ to determine the number of the segment which has generated the pulse.

Three statements are formulated below that make the construction of an optimal single-tact search algorithm an almost obvious procedure.

***Statement 1***. That the procedure of single-tact search was able to define unambiguously the number of the segment $j$, where the point pulse-initiator source is located, it is necessary and sufficient that all possible realizations of the vector $\mathbf{r}$, characterizing the receiving system's state at the moment of pulse registration, differ one from another.

***Statement 2***. The maximal number of elementary segments $N_{max}$ initial search interval $(0,L)$ may be divided into (this number $N_{max}$, actually sets the localization accuracy of the pulsed source), is $N_{max}=2^n-1$, where $n$ is the number of receivers used (the total number of different states of the vector $r$ is $2^n$, but the system can not be in the state $r=0$ when registering the pulse. If we try to exceed the number $N_{max}=2^n-1$ it will become impossible to restore unambiguously the number of a segment with a pulsed source).

***Statement 3***. It is expedient to carry out creation of the binary table $(x_{ij})$ corresponding to the optimal algorithm for the single-tact random point source localization, not by combining row vectors $x_i=(x_{i1},x_{i2},…,x_{iN})$, $i=\overline{1,n}$, describing each of the $n$ receivers, but by combining the column vector $x_j^T=(x_{1j},x_{2j},…,x_{nj})^T$, $i=\overline{1,N}$, characterizing the receiving system's dynamic response to the pulse in the segment $j$.

Considering statements 1-3, the optimal formation of the table $(x_{ij})$ is to choose one of $(2^n-1)!$ variants to pack two-dimensional array that includes all nonzero realization of column vector $\mathbf{x}_j^T=(x_{1j},x_{2j},…,x_{nj})^T$. Each binary column $\{x_{1j},x_{2j},…,x_{nj}\}^T$ corresponds to its number $j=\sum_{i=1}^{n}x_{ij}\times 2^{n-i}$ from the range $j=\overline{1,2^n-1}$. Note that the order of the columns $\mathbf{x}_j^T$ in the formed array $(x_{ij})$ can be any. If the columns of array follow in the increasing order (by their "content"), then the matrix $(x_{ij})$ describing such a receiving system, is presented by the Table 3. In this case, the algorithm to find the segment number with an unknown source is very simple: if the dynamic state of the receiving system is described by the vector $\mathbf{r}=(r_1,r_2,…,r_n)^T$, then the pulsed source is located in the segment with the number $j=\sum_{i=1}^{n}r_i\times 2^{n-i}$, and the set of scalar coordinates of the column vector $\mathbf{r}$ is the binary record of this number.

The absolute accuracy of the single-tact localization procedure of an unknown source is $\varepsilon=L/(2^n-1)$, and the average localization time $<\tau>$ is equal to expected value of pause between the pulses

$$<\tau>=\int_0^\infty th(t)dt=\int_0^\infty t\lambda\exp(-\lambda t)dt=\frac{1}{\lambda}$$

and does not depend on the number of receivers used.

**Table 3. A binary table corresponding to the receiving system with monotonically increasing numbers of pulse-initiator segments.**

| RECEIVERS | Segments | | | | |
|---|---|---|---|---|---|
| | 1 | 2 | 3 | … | $N_{max}=2^n-1$ |
| 1-th receiver | 0 | 0 | 0 | … | 1 |
| 2-th receiver | 0 | 0 | 0 | … | 1 |
| 3-th receiver | 0 | 0 | 0 | … | 1 |
| ⋮ | ⋮ | ⋮ | ⋮ | ⋮ | ⋮ |
| $i$-th receiver | 0 | 0 | 0 | … | 1 |
| ⋮ | ⋮ | ⋮ | ⋮ | ⋮ | ⋮ |
| $(n-1)$-th receiver | | 1 | 1 | … | 1 |
| $n$-th receiver | 1 | 0 | 1 | … | 1 |

*(b) Optimal multi-step search using n receivers*. The single-tact search procedure described above shows how the search effort should be distributed among the $n$ receivers if it is required to reach the highest accuracy of random pulsed source localization. The main moment which will be directly used further is that the procedure described above is extremely constructive and completely unambiguous, so in the case of an optimal distribution of search efforts, the pulsed object is always localized with accuracy $W/(2^n-1)$, where *W* is the integrated viewing window uniting all *n* receivers used. Now, taking these considerations into account, we can begin to solve a more complicated task: it is necessary to construct time-optimal algorithm when the search of a random pulsed source has to be carried out by a system of *n* receiving devices with the required localization accuracy $\varepsilon$. Except stipulated in advance localization accuracy, an important complicating factor in the formulated problem is that the choice of the optimum search procedure isn't limited only by single-tact localization algorithms, and there is no requirement to finish the search at the time of the first pulse generation. Moreover, as will be shown below, even at rather low requirements to the localization accuracy, the optimal algorithm is a multi-stage procedure (transition from one stage to other stage comes at the moment when the receiving system registers the next pulse). At the same time, it is quite acceptable that a part of the pulses generated by a random source is not fixed by the system. As for the previous problem, it is assumed that the localized point source has a uniform distribution density on the interval (0,*L*).

Passing to the solution of the task, let us introduce a number of additional designations. The symbol *M* will be the number of stage in the search procedure, and the symbol $W_i$ will be the aggregate system viewing window uniting *n* receivers at the *i*-th stage of the search (in cases that do not allow double interpretation, the same symbol will mean not only the window itself, but also its linear dimension). Taking this into account, the average time $\langle \tau \rangle$ of any *M*-stage (not necessarily optimal) search procedure of a random source that guarantees the localization accuracy $\varepsilon$ will be

$$\langle \tau \rangle = \frac{1}{\lambda} \times \left( \frac{L}{W_1} + \frac{\frac{W_1}{2^n - 1}}{W_2} + \frac{\frac{W_2}{2^n - 1}}{W_3} + ... + \frac{\frac{W_{M-1}}{2^n - 1}}{W_M} \right). \tag{30}$$

$$W_M = (2^n - 1)\varepsilon. \tag{30a}$$

The integer parameter *M* and continuous variables $W_i$, $i = \overline{1, M}$, which deliver the minimum to the expression (30), need to be estimated. It is taken into account in relation (30) that each subsequent (*i*+1)-th stage of the search procedure includes scanning by the cumulative viewing window $W_{i+1}$ within one of the segments ($2^n$-1) that formed the viewing window $W_i$ on the previous stage, and (*i*+1)-th scanning stage is carried out inside of the segment where *i*-th pulse was fixed. For a fixed *M*, the optimal sizes of scanning viewing windows $W_i$, $i = \overline{1, M}$, for which average localization time (30) reaches a minimum, have to satisfy the system of equations

$$\begin{cases} \frac{\partial \langle \tau \rangle}{\partial W_1} = \frac{1}{\lambda(2^n - 1)} \times \left( -\frac{W_0}{W_1^2} + \frac{1}{W_2} \right) = 0; \\ \frac{\partial \langle \tau \rangle}{\partial W_2} = \frac{1}{\lambda(2^n - 1)} \times \left( -\frac{W_1}{W_2^2} + \frac{1}{W_3} \right) = 0; \\ \vdots \\ \frac{\partial \langle \tau \rangle}{\partial W_{M-1}} = \frac{1}{\lambda(2^n - 1)} \times \left( -\frac{W_{M-2}}{W_{M-1}^2} + \frac{1}{W_M} \right) = 0. \end{cases} \tag{31}$$

$$W_M = (2^n - 1)\varepsilon. \tag{31a}$$

$$W_0 = (2^n - 1)L. \tag{31b}$$

The system of equations (31) is obtained by simple equating to zero all the partial derivatives of average search time $\langle \tau \rangle$ (30), and the relation (31b) is introduced for the symmetrization of the notation. It is more convenient to break the further solution of the problem into two parts. First, we find the number $M_{opt}$ of stages for the optimal localization procedure, and then we calculate all other parameters. Let us note that for a fixed value of the integer parameter *M* the solution of the system (31) is

$$W_i = \left[(2^n - 1)(\varepsilon/L)^{\frac{i}{M}}\right] \times L, \quad i = \overline{1, M}. \tag{32}$$

Taking into account that the dimensionless coefficients $(2^n - 1)(\varepsilon/L)^{\frac{i}{M}}$ can not be more then 1, we get the condition under which the solution (3) is valid:

$$(\varepsilon/L) \leq \frac{1}{(2^n - 1)^M}. \tag{32a}$$

Substituting (32) into (30), we obtain the expression for the average time $<\tau>$ in the case of the $M$-stage localization procedure:

$$\langle \tau \rangle_M = \frac{1}{\lambda} \times \frac{M}{2^n - 1} (\varepsilon/L)^{-\frac{1}{M}}. \tag{33}$$

Function

$$f(x) = \frac{x(\varepsilon/L)^{-\frac{1}{x}}}{\lambda(2^n - 1)}, \tag{34}$$

which is a continuous analogue of expression (33), has only one local minimum (this will be used later), reached at the point $x_{\min} = -\ln(\varepsilon/L)$. Rewriting the condition (32a) in the equivalent form

$$M \leq -\frac{\ln(\varepsilon/L)}{\ln(2^n - 1)}, \tag{35}$$

we find that for any value $n \geq 2$ (positive and integer), the number of $M_{opt}$ stages in the optimal localization procedure is the same as the maximum integer satisfying the constraint (35). Here we use the fact that the function (34) more to the left of a point $x_{\min} = -\ln(\varepsilon/L)$ monotonously decreases. Thus, for the values $\varepsilon/L$ in the neighborhood of the point

$$(\varepsilon/L) = \frac{1}{(2^n - 1)^M}$$

the optimal search procedure consists of $M$ stages. For a complete description of the optimal localization algorithm it remains to find out under what requirements to the accuracy of localization (i.e. for what the values of $\varepsilon/L$) transition from $M$-stage to $(M+1)$-stage search occurs. It is obvious that at the transition point the average time of the $M$-stage search has to be equal to the average time of the most "fast" $(M+1)$-stage search:

$$\frac{1}{\lambda} \times \frac{M}{2^n - 1} (\varepsilon/L)^{-\frac{1}{M}} = \frac{M+1}{\lambda}. \tag{36}$$

From here we obtain the point where the optimal localization algorithms have a transition from the $M$-stage search strategy to $(M+1)$-stage one:

$$(\varepsilon/L)_{M \to M+1} = \frac{1}{(2^n - 1)^M} \times \left(\frac{M}{M+1}\right)^M. \tag{37}$$

The main calculation results that systematize the parameters of time-optimal localization algorithms of objects that in random timepoints generate instantaneous pulses (delta functions) are presented in Table 4. These results are applicable to systems with several ($n \geq 2$) receivers. The last line of the table contains the parameters of the optimal search for the asymptotic procedure, when the required localization accuracy $\varepsilon/L$ tends to zero.

**Table 4. Parameters of the optimal search for the random pulsed source depending on the number of receivers $n$ ($n \geq 2$) and the required localization accuracy $\varepsilon$.**

| $(\varepsilon/L)$ (required localization accuracy) | $M_{opt}$ * | $W_m, m = \overline{1, M_{opt}}$ (Viewing windows of the receiving system at each of $M_{opt}$ stages of optimal search) | $\langle \tau \rangle$ (average localization time) |
|---|---|---|---|
| $\dfrac{1}{2^n - 1} \leq (\varepsilon/L) < 1$ | 1 | $W_1 = L$ | $\dfrac{1}{\lambda}$ |
| $\dfrac{1}{2(2^n - 1)} \leq (\varepsilon/L) \leq \dfrac{1}{2^n - 1}$ | 1 | $W_1 = (2^n - 1)(\varepsilon/L)^1 \times L = (2^n - 1)\varepsilon$ | $\dfrac{1}{\lambda(2^n - 1)}(\varepsilon/L)^{-1}$ |
| $\dfrac{1}{(2^n - 1)^2} \leq (\varepsilon/L) \leq \dfrac{1}{2(2^n - 1)}$ | 2 | $W_1 = L$ <br> $W_2 = \dfrac{1}{2^n - 1} \times L$ | $\dfrac{2}{\lambda}$ |
| $\dfrac{1}{(2^n - 1)^2} \times \left(\dfrac{2}{3}\right)^2 \leq$ <br> $\leq (\varepsilon/L) \leq \dfrac{1}{(2^n - 1)^2}$ | 2 | $W_1 = (2^n - 1)(\varepsilon/L)^{\frac{1}{2}} \times L$ <br> $W_2 = (2^n - 1)(\varepsilon/L) \times L = (2^n - 1)\varepsilon$ | $\dfrac{2}{\lambda(2^n - 1)}(\varepsilon/L)^{-\frac{1}{2}}$ |
| $\dfrac{1}{(2^n - 1)^3} \leq (\varepsilon/L) \leq$ <br> $\leq \dfrac{1}{(2^n - 1)^2} \times \left(\dfrac{2}{3}\right)^2$ | 3 | $W_1 = L$ <br> $W_2 = \dfrac{1}{2^n - 1} \times L$ <br> $W_3 = \dfrac{1}{(2^n - 1)^2} \times L$ | $\dfrac{3}{\lambda}$ |
| $\dfrac{1}{(2^n - 1)^3} \times \left(\dfrac{3}{4}\right)^3 \leq$ <br> $\leq (\varepsilon/L) \leq \dfrac{1}{(2^n - 1)^3}$ | 3 | $W_1 = (2^n - 1)(\varepsilon/L)^{\frac{1}{3}} \times L$ <br> $W_2 = (2^n - 1)(\varepsilon/L)^{\frac{2}{3}} \times L$ <br> $W_3 = (2^n - 1)(\varepsilon/L) \times L = (2^n - 1)\varepsilon$ | $\dfrac{3}{\lambda(2^n - 1)}(\varepsilon/L)^{-\frac{1}{3}}$ |
| $\vdots$ | $\vdots$ | $\vdots$ | $\vdots$ |
| $\dfrac{1}{(2^n - 1)^M} \leq (\varepsilon/L) \leq$ <br> $\leq \dfrac{1}{(2^n - 1)^{M-1}} \left(\dfrac{M-1}{M}\right)^{M-1}$ | $M$ | $W_1 = L$ <br> $W_2 = \dfrac{1}{2^n - 1} \times L$ <br> ... <br> $W_m = \dfrac{1}{(2^n - 1)^{m-1}} \times L$ <br> ... <br> $W_M = \dfrac{1}{(2^n - 1)^{M-1}} \times L$ | $\dfrac{M}{\lambda}$ |

| $(\varepsilon/L)$ | $M$ | $W_m, m=\overline{1,M}$ | $\langle\tau\rangle$ |
|---|---|---|---|
| $\frac{1}{(2^n-1)^M}\left(\frac{M}{M+1}\right)^M \leq (\varepsilon/L) \leq$ $\leq \frac{1}{(2^n-1)^M}$ | $M$ | $W_1 = (2^n-1)(\varepsilon/L)^{\frac{1}{M}} \times L$ $W_2 = (2^n-1)(\varepsilon/L)^{\frac{2}{M}} \times L$ $\vdots$ $W_m = (2^n-1)(\varepsilon/L)^{\frac{m}{M}} \times L$ $\vdots$ $W_M = (2^n-1)(\varepsilon/L)^{\frac{M}{M}} \times L = (2^n-1)\varepsilon$ | $\frac{M}{\lambda(2^n-1)}(\varepsilon/L)^{-\frac{1}{M}}$ |
| $\vdots$ | $\vdots$ | $\vdots$ | $\vdots$ |
| $(\varepsilon/L) \to 0$ $\frac{e^{-1}}{(2^n-1)^{M_\infty}} \leq (\varepsilon/L) \leq$ $\leq \frac{e^{-1}}{(2^n-1)^{M_\infty-1}}$ | $M_\infty \approx$ $\approx \frac{-\ln(\varepsilon/L)}{\ln(2^n-1)}$ | $W_1 = L$ $W_2 = \frac{1}{2^n-1} \times L$ ... $W_m = \frac{1}{(2^n-1)^{m-1}} \times L$ ... $W_{M_\infty} = \frac{1}{(2^n-1)^{M_\infty-1}} \times L \approx$ $\approx \frac{2^n-1}{(2^n-1)^{\frac{-\ln(\varepsilon/L)}{\ln(2^n-1)}}} \times L = (2^n-1)\varepsilon$ | $\frac{M_\infty}{\lambda} \approx$ $\approx -\frac{\ln(\varepsilon/L)}{\lambda\ln(2^n-1)}$ |

\* Optimal number of stages for given localization accuracy

**Table 5.** Optimal search parameters of a random pulsed source for a system with one receiver ($n = 1$).

| $(\varepsilon/L)$ (required localization accuracy) | $M_{opt}$ * | $W_m, m=\overline{1,M_{opt}}$ (Viewing windows of the receiving system for each of $M_{opt}$ stages of optimal search) | $\langle\tau\rangle$ (average localization time) |
|---|---|---|---|
| $\frac{1}{4} \leq (\varepsilon/L) < 1$ | 1 | $W_1 = \varepsilon$ | $\frac{1}{\lambda}(\varepsilon/L)^{-1}$ |
| $\left(\frac{2}{3}\right)^6 \leq (\varepsilon/L) \leq \frac{1}{4}$ | 2 | $W_1 = (\varepsilon/L)^{\frac{1}{2}} \times L$ $W_2 = (\varepsilon/L) \times L = \varepsilon$ | $\frac{2}{\lambda}(\varepsilon/L)^{-\frac{1}{2}}$ |
| $\left(\frac{3}{4}\right)^{12} \leq (\varepsilon/L) \leq \left(\frac{2}{3}\right)^6$ | 3 | $W_1 = (\varepsilon/L)^{\frac{1}{3}} \times L$ $W_2 = (\varepsilon/L)^{\frac{2}{3}} \times L$ $W_3 = (\varepsilon/L) \times L = \varepsilon$ | $\frac{3}{\lambda}(\varepsilon/L)^{-\frac{1}{3}}$ |
| $\vdots$ | $\vdots$ | $\vdots$ | $\vdots$ |
| $\left(\frac{M}{M+1}\right)^{M(M+1)} \leq (\varepsilon/L) \leq$ $\leq \left(\frac{M-1}{M}\right)^{M(M-1)}$ | $M$ | $W_1 = (\varepsilon/L)^{\frac{1}{M}} \times L$ $W_2 = (\varepsilon/L)^{\frac{2}{M}} \times L$ $\vdots$ $W_m = (\varepsilon/L)^{\frac{m}{M}} \times L$ $\vdots$ $W_M = (\varepsilon/L)^{\frac{M}{M}} \times L = \varepsilon$ | $\frac{M}{\lambda}(\varepsilon/L)^{-\frac{1}{M}}$ |
| $\vdots$ | $\vdots$ | $\vdots$ | $\vdots$ |

| $(\varepsilon/L) \to 0$ $(\varepsilon/L) \approx e^{-M_\infty}$ | $M_\infty \approx$ $\approx -\ln(\varepsilon/L)$ | $W_1 = (\varepsilon/L)^{\frac{1}{M_\infty}} \times L = e^{-1} \times L$ $W_2 = (\varepsilon/L)^{\frac{2}{M_\infty}} \times L = e^{-2} \times L$ $\vdots$ $W_m = (\varepsilon/L)^{\frac{m}{M_\infty}} \times L = e^{-m} \times L$ $\vdots$ $W_{M_\infty} = e^{-M_\infty} \times L = (\varepsilon/L) \times L = \varepsilon$ | $\frac{M_\infty}{\lambda}(\varepsilon/L)^{\frac{1}{M_\infty}} \approx$ $\approx \frac{-\ln(\varepsilon/L)}{\lambda}(\varepsilon/L)^{\frac{1}{\ln(\varepsilon/L)}} =$ $= -\frac{e\ln(\varepsilon/L)}{\lambda}$ |
|---|---|---|---|

\* Optimal number of stages for given localization accuracy

**Conclusion**

The obtained results solve the problem of creation the algorithms of time-optimal localization of random pulsed point sources, when such objects have a uniform distribution on the search interval (0,*L*). The proposed search strategies open up the prospect to find the optimal localization algorithms in cases when the probability density function of random pulsed source differs from uniform. Another interesting and poorly investigated direction of the problem is the construction of optimal search procedures focused on the simultaneous localization of several random sources. The described algorithms are partially presented in [5,6].

The authors are grateful to the Russian Foundation for Basic Research for the support (grants №№ 16-01-00313, 13-01-00361) of scientific works, which are the basis for this publication.

*Bibliography*

1. Potapov A.A., Sokolov A.V. Perspective methods of processing of radar signals on the basis of fractal and textural measures. *Izvestiya of RAS. Series Physical, 2003, V. 67, №12, p. 1775-1778 (in Russian)*
2. Howland, P.E.: "A Passive Metric Radar Using the Transmitters of Opportunity", *Int. Conf. on Radar, Paris, France, May 1994, p. 251–256.*
3. Mark R. Bell, "Information theory and radar waveform design." *IEEE Transactions on Information Theory, 39.5 (1993): p. 1578–1597.*
4. V.M. Efimov, A.A. Nesterov, A.L. Reznik. Algorithms of the optimal search for light point objects in speed. *Avtometriya, 1980, №3, p.72-76 (in Russian).*
5. A.L. Reznik. The programs for analytical calculations in the problems of point object localization. *Optoelectronics, instrumentation and data processing, 1991, №6, p.21-24.*
6. A.L. Reznik, A.V. Tuzikov, A.A. Soloviev, A.V. Torgov. Time-Optimal Algorithms of Searching for Pulsed-Point Sources for Systems with Several Detectors. *Optoelectronics, instrumentation and data processing, 2017, №3, p. 3-11.*